\documentclass[%
 aip,
 amsmath,amssymb,
 reprint,%
]{revtex4-1}
\usepackage{graphicx}
\usepackage{dcolumn}
\usepackage{bm}

\usepackage[utf8]{inputenc}
\usepackage[T1]{fontenc}
\usepackage{mathptmx}
\usepackage{etoolbox}
\usepackage{listings}
\usepackage[colorlinks=true,
            linkcolor=blue,
            urlcolor=blue,
            citecolor=blue]{hyperref}
\usepackage{color}
\usepackage{graphics}
\usepackage{tikz}
\usepackage{soul}

\usetikzlibrary{arrows,shapes,positioning,shadows,trees}

\newcommand{\nl}{\nonumber \\}
\DeclareUnicodeCharacter{0301}{\'{i}}
\DeclareUnicodeCharacter{0308}{\"{0}}
\definecolor{dkgreen}{rgb}{0,0.6,0}
\definecolor{gray}{rgb}{0.5,0.5,0.5}
\definecolor{mauve}{rgb}{0.58,0,0.82}

\makeatletter
\def\@email#1#2{%
 \endgroup
 \patchcmd{\titleblock@produce}
  {\frontmatter@RRAPformat}
  {\frontmatter@RRAPformat{\produce@RRAP{*#1\href{mailto:#2}{#2}}}\frontmatter@RRAPformat}
  {}{}
}%
\makeatother
\begin{document}

\preprint{AIP/123-QED}

\title[]{\textsc{MLQD}: A package for machine learning-based quantum dissipative dynamics}
\author{Arif Ullah}
\affiliation{School of Physics and Optoelectronic Engineering, Anhui University, Hefei 230601, China}
\author{Pavlo O. Dral}%
 \email{dral@xmu.edu.cn}
 \email{arif@ahu.edu.cn}
\affiliation{ 
State Key Laboratory of Physical Chemistry of Solid Surfaces, College of Chemistry and Chemical Engineering, Fujian Provincial Key Laboratory of Theoretical and Computational Chemistry, and Innovation Laboratory for Sciences and Technologies of Energy Materials of Fujian Province (IKKEM), Xiamen University, Xiamen 361005, China
}%

\date{\today}
\begin{abstract}
Machine learning has emerged as a promising paradigm to study the quantum dissipative dynamics of open quantum systems. To facilitate the use of our recently published ML-based approaches for quantum dissipative dynamics, here we present an open-source Python package \textsc{MLQD} (\url{https://github.com/Arif-PhyChem/MLQD}), which currently supports the three ML-based quantum dynamics approaches: (1) the recursive dynamics with kernel ridge regression (KRR) method, (2) the non-recursive artificial-intelligence-based quantum dynamics (AIQD) approach and (3) the blazingly fast one-shot trajectory learning (OSTL) approach, where both AIQD and OSTL use the convolutional neural networks (CNN). This paper describes the features of the \textsc{MLQD} package, the technical details, optimization of hyperparameters, visualization of results, and the demonstration of the \textsc{MLQD}'s applicability for two widely studied systems, namely the spin-boson model and the Fenna--Matthews--Olson (FMO) complex. To make \textsc{MLQD} more user-friendly and accessible, we have made it available on the Python Package Index (PyPi) platform and it can be installed via \lstinline{pip install mlqd}. In addition, it is also available on the XACS cloud computing platform (\url{https://XACScloud.com}) via the interface to the \textsc{MLatom} package (\url{http://MLatom.com}).  \\
\textbf{Program summary} \\
\textit{Program Title}: MLQD \\
\textit{Developer's repository link}: \url{https://github.com/Arif-PhyChem/MLQD} \\
\textit{Code Ocean capsule}: \url{https://codeocean.com/capsule/5563143/tree} \\
\textit{Licensing provisions}: Apache Software License 2.0 \\
\textit{Programming language}: Python 3.0 \\
\textit{Supplementary material}: Jupyter Notebook-based tutorials \\
\textit{External routines/libraries:}: Tensorflow, Scikit-learn, Hyperopt, Matplotlib, MLatom \\
\textit{Nature of problem}: Fast propagation of quantum dissipative dynamics with machine learning approaches. \\
\textit{Solution method}: We have developed MLQD as a comprehensive framework that streamlines and supports the implementation of our recently published machine learning-based approaches for efficient propagation of quantum dissipative dynamics. This framework encompasses: (1) the recursive dynamics with kernel ridge regression (KRR) method, as well as the non-recursive approaches utilizing convolutional neural networks (CNN), namely (2) artificial intelligence-based quantum dynamics (AIQD), and (3) one-shot trajectory learning (OSTL). \\
\textit{Unusual or notable features}:  
\begin{enumerate}
 \vspace{-0.2cm}
    \item Users can train a machine learning (ML) model following one of the ML-based approaches: KRR, AIQD and OSTL.  
    \vspace{-0.3cm}\item Users have the option to propagate dynamics with the existing trained ML models.
    \vspace{-0.3cm}\item MLQD also provides the transformation of trajectories into the training data.
    \vspace{-0.3cm}\item MLQD also supports hyperparameter optimization using MLATOM’s grid search functionality for KRR and Bayesian methods with Tree-structured Parzen Estimator (TPE) for CNN models via the HYPEROPT package.
    \vspace{-0.2cm}\item MLQD also facilitates the visualization of results via auto-plotting.
    \vspace{-0.2cm}\item MLQD is designed to be user-friendly and easily accessible, with availability on the XACS cloud computing platform (https://XACScloud.com) via the interface to the MLATOM package (http://MLatom.com). In addition, it is also available as a pip package which makes it easy to install. 
\end{enumerate}
\textit{Future outlook}: MLQD will be extended to more realistic systems along with the incorporation of other machine learning-based approaches as well as the traditional quantum dynamics methods.

\end{abstract}

\maketitle
\onecolumngrid
\section{\label{sec:intro}Introduction}
The basic time-dependent Schr{\"{o}}dinger equation describes the unitary dynamics of an isolated quantum system. However, isolated quantum systems are an idealistic approximation with many limitations as in real world, systems are always coupled to an environment. Thus, to study quantum systems in reality, it is important to incorporate the effects of environment which gives rise to dephasing and dissipation. Systems with dephasing and dissipation (open quantum systems or quantum dissipative systems) are ubiquitous and can be exploited in environment-assisted quantum transport,\cite{plenio2008dephasing, rebentrost2009environment} chemical and biological systems,\cite{may2008charge, dorfman2013photosynthetic,slocombe2022open} quantum information processing and quantum computing,\cite{verstraete2009quantum, unruh1995maintaining} defect tunneling in solids,\cite{golding1992dissipative,vojta2001kondo} quantum electrodynamics,\cite{wen2004quantum,le2018driven} colour centres and Cooper pair boxes,\cite{reed2012realization,georgescu2020trapped} quantum optics,\cite{carmichael2009open, sieberer2016keldysh} superconducting junctions,\cite{milovsevic2010ginzburg} and quarkonium transport in a hot nuclear environment.\cite{yao2021open}  Exact solution of Schr{\"{o}}dinger equation for open quantum systems is a daunting task and in most cases is not feasible because of exponential growth in Hilbert space dimension and a large number of environment degrees of freedom. Thus, many approximations are adopted such as averaging out environment degrees of freedom\cite{wang2022quantum} and classical description of the system and/or environment.\cite{jain2022pedagogical, meyer1979classical}

In the past three decades, great progress has been made in the development of theoretical approaches for open quantum systems. These approaches can be categorized into three distinct categories: (1) the fully classical approaches \cite{meyer1979classical,stock1997semiclassical, runeson2019spin, liu2021unified} where system and environment both are described on the same classical footings, (2) the quantum-classical approaches \cite{barbatti2011nonadiabatic} where system is described quantum mechanically while environment is treated classically, and (3) the fully quantum approaches which treat both system and environment quantum mechanically. The last category can be further divided into Markovian approaches such as perturbative Redfield equation \cite{redfield1957theory} and a long list of non-Markovian approaches such as the multi-configuration time-dependent Hartree (MCTDH),\cite{beck2000multiconfiguration, meyer2009multidimensional} the exact Nakajima--Zwanzig formalism\cite{nakajima1958quantum,zwanzig1960ensemble} and its kernel-based expansions,\cite{silbey1984variational, leggett1987dynamics, xu2018convergence}  Green's function formalism,\cite{bergmann2021green} the transfer tensor method\cite{cerrillo2014non} and its extension,\cite{kananenka2016accurate}  the pseudo-mode approach,\cite{garraway1997nonperturbative, luo2023quantum} the reaction coordinate (RC) approach,\cite{nazir2018reaction,anto2021capturing} the quantum Monte Carlo (QMC),\cite{schiro2009real,cohen2011memory} the time-dependent numerical renormalization group (NRG),\cite{anders2005real} the density matrix renormalization group (tDMRG),\cite{schollwock2005density} the polaron transformation,\cite{silbey1984variational,xu2016non} the time evolving density matrix using the orthogonal polynomial algorithm (TEDOPA),\cite{chin2010exact, prior2010efficient} the quasiadiabatic propagator path integral (QuAPI),\cite{makarov1994path,makri1995numerical} the numerical variational method (NVM),\cite{zhou2018variational} the automated compression of environment (ACE) method,\cite{cygorek2021numerically} the hierarchy equations of motion (HEOM),\cite{tanimura1989time, yan2004hierarchical,tanimura2006stochastic,jin2008exact,shi2009efficient,hu2010communication,moix2013hybrid,liu2014reduced,gong2018quantum,han2018exact,cui2019highly,zhang2020hierarchical,chen2022universal} the dissipation equation of motion (DEOM),\cite{xu2015dissipaton,ying2023spin} and the stochastic equation of motion (SEOM).\cite{stockburger2002exact,koch2008non,stockburger2016exact,schmitz2019variance,shao2004decoupling,shao2010rigorous, hsieh2018unified, mccaul2017partition, ke2016hierarchy, ke2017extension,han2019stochastic,han2020stochastic,ullah2020stochastic,yan2022piecewise,shao2022dynamics} These methods on their own are successful attempts to solve the complex multibody problem of open quantum systems, however, the prohibitive increase in their computational cost with the system size limits their applicability.

In the past two decades, data proliferation has led to the advent of machine learning (ML) methods. ML has been described as a fourth pillar in Science next to experiment, theory and simulation.\cite{von2020introducing}  Without discussing the broad use of ML, in recent years, ML has seen a surge in the field of quantum dynamics in general and, in particular, quantum dissipative dynamics.\cite{zhang2023excited, choi2022learning, kadupitiya2022solving,luchnikov2022probing,yao2022emulating, wang2021spacetime,banchi2018modelling,  hartmann2019neural, ge2022four, ullah2021speeding, rodriguez2022comparative, herrera2021convolutional, wu2021forecasting, lin2022automatic, bandyopadhyay2018applications, yang2020applications, ullah2022predicting, ullah2022one, hase2016machine,naicker2022machine, hase2017machine,farahvash2020machine,secor2021artificial,akimov2021extending,lin2022trajectory,tang2022fewest} To be specific, ML has been used to predict molecular configurations in the four-dimensional space,\cite{ge2022four} the relaxation dynamics of a two-state system\cite{ullah2021speeding, rodriguez2022comparative, herrera2021convolutional, wu2021forecasting, lin2022automatic, bandyopadhyay2018applications, yang2020applications} the excitation energy transfer in Fenna--Matthews--Olson light-harvesting complex,\cite{ullah2022predicting, ullah2022one, hase2016machine,naicker2022machine} average exciton transfer times and transfer efficiencies,\cite{hase2017machine} parameters of Hamiltonian,\cite{farahvash2020machine} evolution of the proton density in a potential well\cite{secor2021artificial} and vibronic Hamiltonians as a direct function of time.\cite{akimov2021extending} Machine learning has also been extended to Meyer--Miller mapping based symmetrical quasi-classical\cite{lin2022trajectory} and fewest-switches surface hopping dynamics.\cite{tang2022fewest}

The rapid development of new ML methods for quantum dissipative dynamics led to so-far not well-organized and scattered software implementations. In this article, we present an open-source software package \textsc{MLQD}, version 1, which provides a framework for ML-based quantum dissipative dynamics implementations. \textsc{MLQD} is incorporated with kernel ridge regression (KRR) and convolutional neural networks (CNN) models and a user can train and predict dynamics following both recursive and non-recursive approaches. We follow our recently published recursive KRR-based approach on the relaxation dynamics of the spin-boson model\cite{ullah2021speeding, rodriguez2022comparative} and non-recursive CNN-based AIQD and OSTL approaches.\cite{ullah2022predicting, ullah2022one} \textsc{MLQD} also supports hyperparameter optimization using \textsc{MLatom}'s grid search functionality\cite{dral2019mlatom, dral2021mlatom, MLatomdev} for KRR and Bayesian methods with Tree-structured Parzen Estimator (TPE)\cite{bergstra2011algorithms} for CNN models via the \textsc{hyperopt} package.\cite{Bergstra2015HyperoptAP} In addition, we also incorporate the visualization of results by auto-plotting. \textsc{MLQD} has also been interfaced with the \textsc{MLatom} package\cite{dral2019mlatom, dral2021mlatom, MLatomdev} which allows the user to run \textsc{MLQD} on the XACS cloud computing platform.\cite{xacs2022}         

In the following, we provide an overview of the \textsc{MLQD} package, the theory of implemented approaches, technical details, optimization of hyperparameters, visualization of results, and the demonstration of \textsc{MLQD}'s applicability for two widely studied systems, namely the spin-boson model and the Fenna--Matthews--Olson (FMO) complex.       

\section{\textsc{MLQD} package overview}
\begin{figure}
    \centering
    \includegraphics[width=\textwidth]{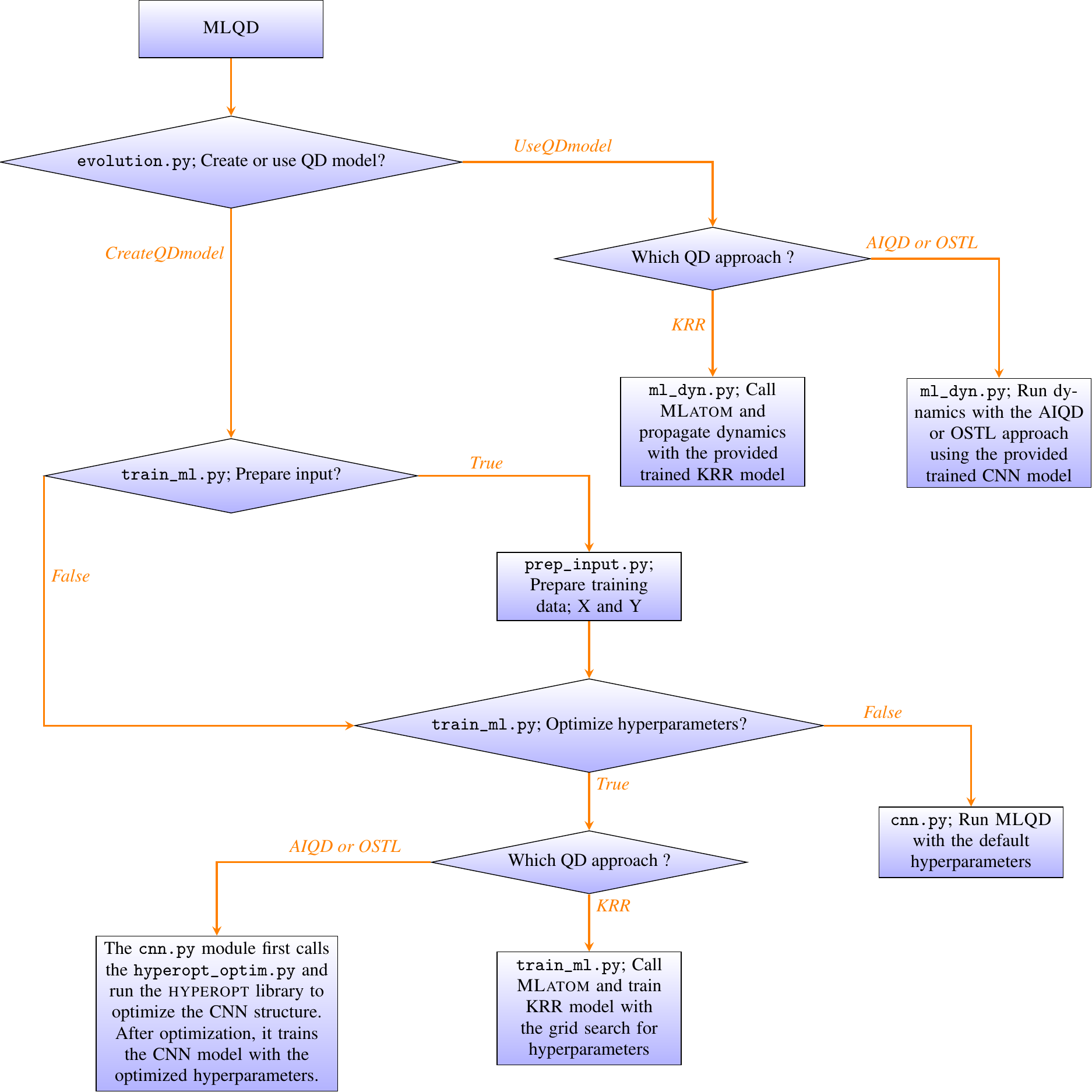}
 \caption{A simplified flowchart of the \textsc{MLQD} package. \textsc{MLQD} package comes with six main python modules: \lstinline{evolution.py}, \lstinline{train_ml.py}, \lstinline{ml_dyn.py}, \lstinline{prep_input.py}, \lstinline{hyperopt_optim.py} and \lstinline{cnn.py}. The \lstinline{evolution.py} module is the headquarter of the MLQD package which controls all the operations including, training a QD model and using the already trained QD model for dynamics propagation--calling \lstinline{train_ml.py} and \lstinline{ml_dyn.py} with the keywords \lstinline{createQDmodel} and \lstinline{useQDmodel}, respectively. Calling \lstinline{train_ml.py} in return calls the \lstinline{cnn.py} and MLatom package\cite{dral2019mlatom, dral2021mlatom, MLatomdev} to train a CNN model and a KRR model, respectively. \textsc{MLQD} has also the feature of preparing the training data X and Y (using \lstinline{prep_input.py}) considering the training trajectories are given in the same format as in our QD3SET-1 database.\cite{ullah2023qd3set} \textsc{MLQD} can also optimize the hyperparameters using the grid search for KRR model (utilizing \textsc{MLatom} in the backend) and \textsc{hyperopt} library (using \lstinline{hyperopt_optim.py})\cite{Bergstra2015HyperoptAP} in the case of AIQD and OSTL approaches. MLQD's architecture allows for future developments with independent modules easily incorporated in the \lstinline{evolution.py} module.}
\label{fig:mlqd}
\end{figure}

\textsc{MLQD} package is written in Python language and provides the implementation of our recently proposed ML-based approaches for quantum dissipative dynamics.\cite{ullah2021speeding,ullah2022predicting, ullah2022one} This section lays out the concise documentation of theory, code design, use and implementation. Coming to the code design, we provide a simplified flowchart of \textsc{MLQD} architecture in Fig.~\ref{fig:mlqd}. \textsc{MLQD} comes with two main features, \lstinline{createQDmodel} which trains a QD model and \lstinline{useQDmodel} which uses the already trained QD model for dynamics propagation. \textsc{MLQD} has also the feature of preparing the training data X and Y considering the training trajectories are given in the same format as in our QD3SET-1 database.\cite{ullah2023qd3set} \textsc{MLQD} can also optimize the hyperparameters using the grid search for KRR model (utilizing \textsc{MLatom}\cite{dral2019mlatom, dral2021mlatom, MLatomdev} in the backend) and \textsc{hyperopt} library\cite{Bergstra2015HyperoptAP} in the case of AIQD and OSTL approaches.  

\subsection{ML-based quantum dynamics approaches}
ML-based quantum dynamics approaches can be divided into two main categories: recursive approaches and non-recursive. 
\subsubsection{Recursive approaches}
\begin{figure}
    \centering
    \includegraphics[width=\textwidth]{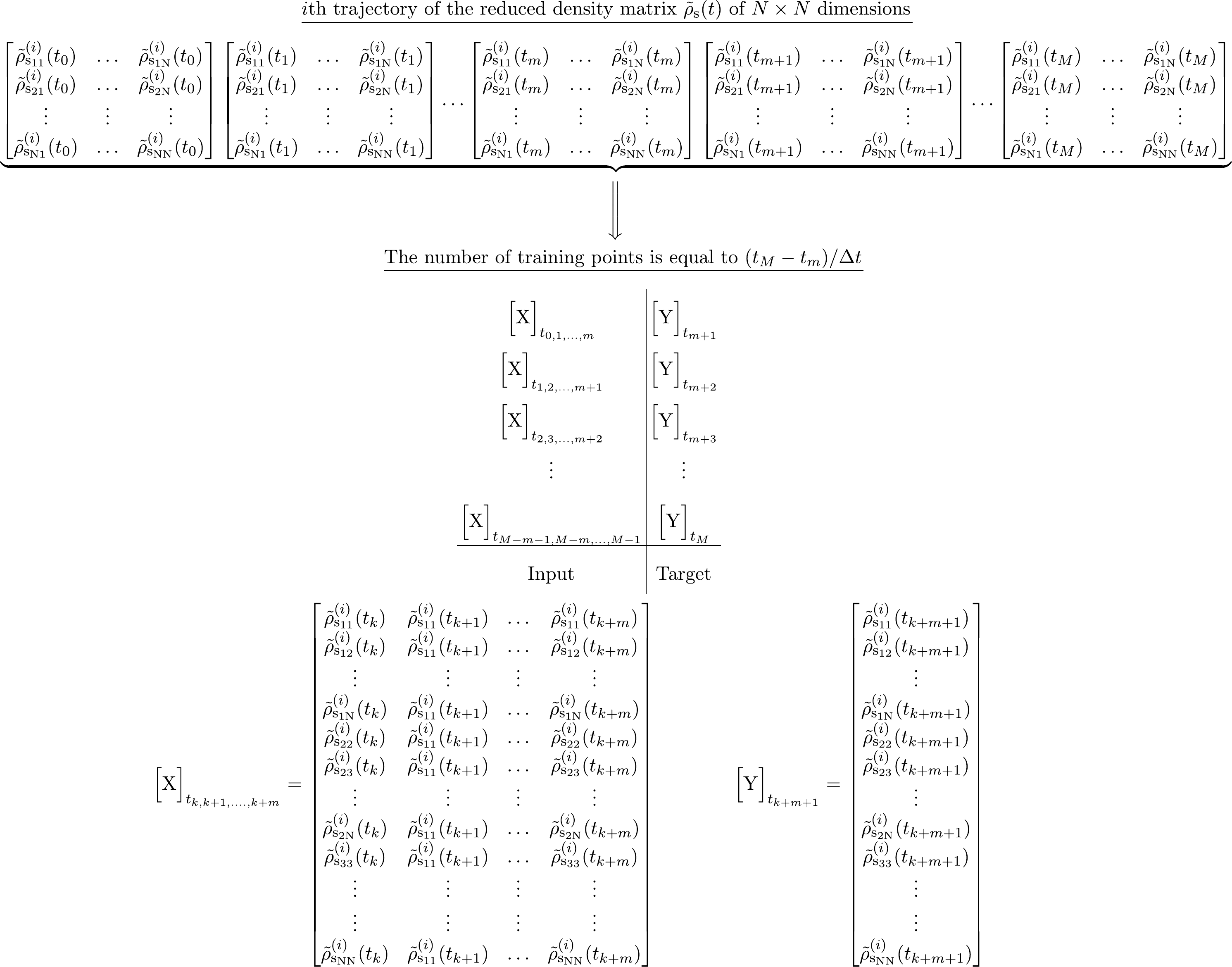}
     \caption{Transformation of the quantum dynamics trajectories into the training data for the recursive approach on an example of the system's RDM $\tilde{\rho}^{(i)}_{\rm s}(t)$.}
    \label{fig:krr}
\end{figure}

In recursive approaches, the future dynamics depends on its past dynamics which in nature is the same as in traditional quantum dynamics approaches. In these approaches, an initial shot-time dynamics of time-length $t_m$ is used as an input to predict system dynamics at the next time step $t_{m+1}$, i.e., $\tilde{\rho}_{\rm s}(t_{m+1}) = f_\text{ML}\left(\tilde{\rho}_{\rm s}(t_{0}), \tilde{\rho}_{\rm s}(t_{2}), \dots, \tilde{\rho}_{\rm s}(t_{m})\right)$ where $\tilde{\rho}_{\rm s}(t)$ is the reduced density matrix (RDM) of the system at time $t$ (see Section~\ref{sec:applications}) and $f_\text{ML}$ is ML function. To incorporate the recursiveness of the dynamics, in the next step, the predicted dynamics at $t_{m+1}$ is appended to the end of the input vector while the value at the start of the input vector is dropped which leads to a new input of the same size as the old input. The new input is used to predict system dynamics at the next time step $t_{m+2}$, i.e., $\tilde{\rho}_{\rm s}(t_{m+2}) = f_\text{ML}\left(\tilde{\rho}_{\rm s}(t_{1}), \tilde{\rho}_{\rm s}(t_{2}), \dots, \tilde{\rho}_{\rm s}(t_{m+1})\right)$. To predict system dynamics at $t_{m+3}$, the predicted dynamics at $t_{m+2}$ is included with the drop of time step at the first end and this process continues till the last time step. In Fig.~\ref{fig:krr}, we have elaborated on the transformation of the training trajectories into the training data for machine learning. Following Ref.~\citenum{ullah2021speeding}, in \textsc{MLQD}, the recursive approach is adopted only for the KRR model and we will refer to this approach as the KRR approach (see Subsection~\ref{subsubsec:krr}).   

\subsubsection{Non-recursive approaches}
\begin{figure}
    \centering
    \includegraphics[width=\textwidth]{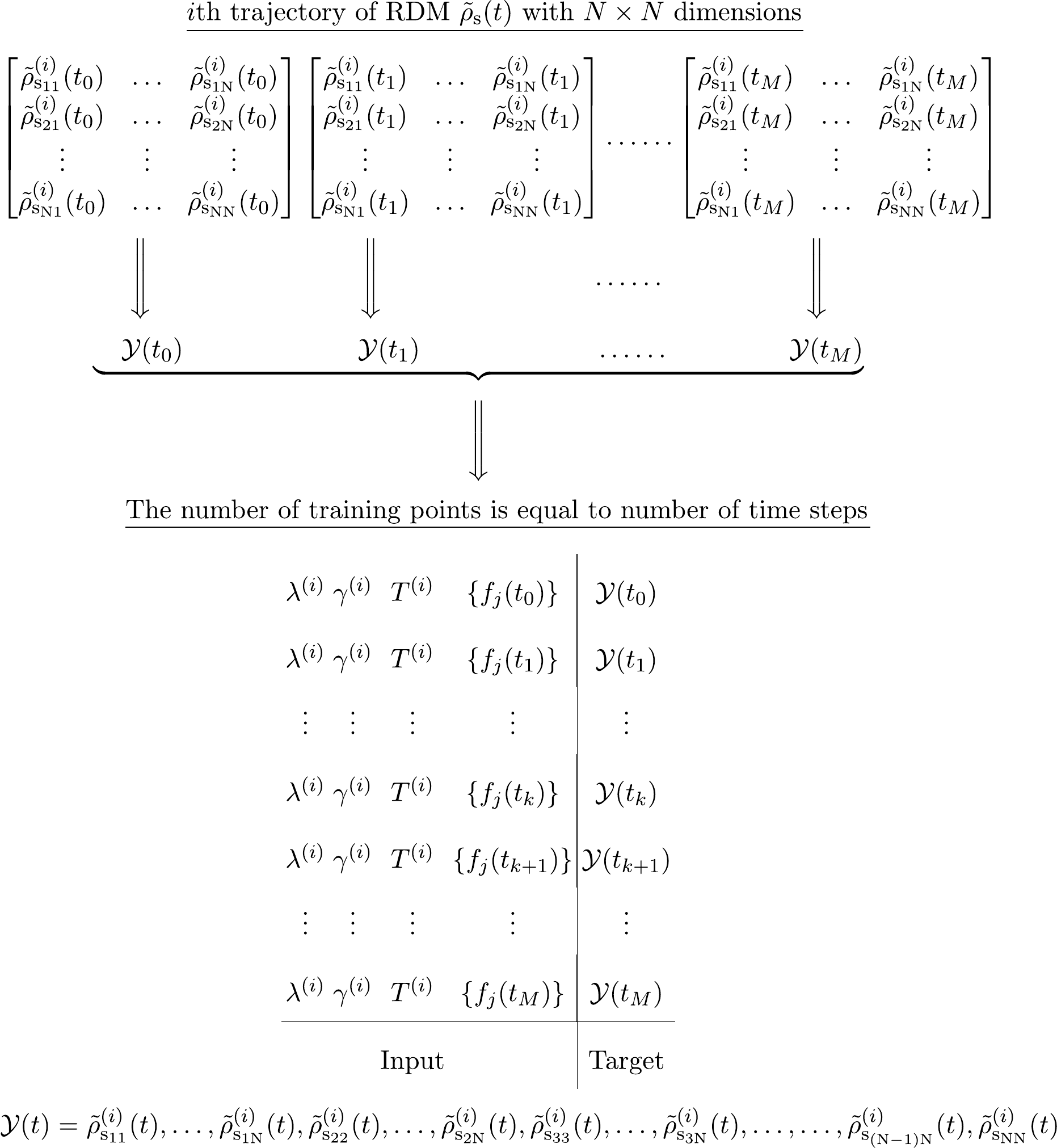}
   \caption{Transformation of the $i$th trajectory of system's RDM $\tilde{\rho}^{(i)}_{\rm s}(t)$ in the AIQD approach. The $\tilde{\rho}^{(i)}_{\rm s}(t)$ at each time step transforms into a vector $\mathbf{\mathcal{Y}}(t)$ with dimension $M$ =  number of sites + (2 $\times$ number of the upper off-diagonal terms). As in RDM $\tilde{\rho}_{\rm s_{nm}}(t) = \tilde{\rho}_{\rm s_{mn}}^*(t) \, (n \neq m)$, only the upper off-diagonal terms are learned. In addition, the real and imaginary parts of each off-diagonal term are separated. Simulation parameters $\lambda^{(i)}$, $\gamma^{(i)}$ and $T^{(i)}$ are the reorganization energy, characteristic frequency, and temperature of the $i$th trajectory in their respective order. The $\{f(t)\}$ is a set of logistic functions normalizing the dimension of time, i.e., $f_j(t) = a/\left(1 + b\exp\left(-(t+c_j)/d\right)\right)$ where $a$, $b$, $d$ are fixed constants while $c_j = 5j-1$ having $j$ as a natural number, i.e., $j\in\{0,1,2,3,\dots\dots\}$.}
    \label{fig:aiqd}
\end{figure}

Recursive approaches have the downside that they should be run sequentially, one step at a time, which is intrinsically computationally costly. An additional downside is the potential for error accumulation at each time step. To develop approaches that are free of these downsides, we have recently proposed two non-recursive approaches AIQD (artificial intelligence-based quantum dynamics\cite{ullah2022predicting}) and OSTL (one-shot trajectory learning \cite{ullah2022one}). Non-recursive approaches are based on neural networks, details are given in Subsection~\ref{subsubsec:nn}. In the following, we briefly describe both approaches.
\paragraph{AIQD approach.}
In the AIQD approach,\cite{ullah2022predicting} system's dynamics (system's RDM) is predicted as a continuous function of time, i.e., dynamics property (system's state) can be predicted at an arbitrary time without step-wise dynamics propagation. Unlike recursive approaches, AIQD does not need to use short-time dynamics as an input. Input, in addition to time, includes simulation parameters such as temperature $T$, characteristic frequency $\gamma$ and system-bath coupling strength $\lambda$. The dynamics corresponding to these parameters is used as a target vector (for training) and predicted by the model when used for inference. In addition, as all time steps are independent of each other, parallel computation of all time steps is possible. Fig.~\ref{fig:aiqd} shows data preparation for the AIQD approach where each trajectory transforms into the training points equal to the number of time steps. We also note that to make predictions in asymptotic limit and cover different time regions with similar accuracy, the time input is mapped into a multi-dimensional vector after normalization with the logistic function.

\begin{figure}
    \centering
    \includegraphics[width=\textwidth]{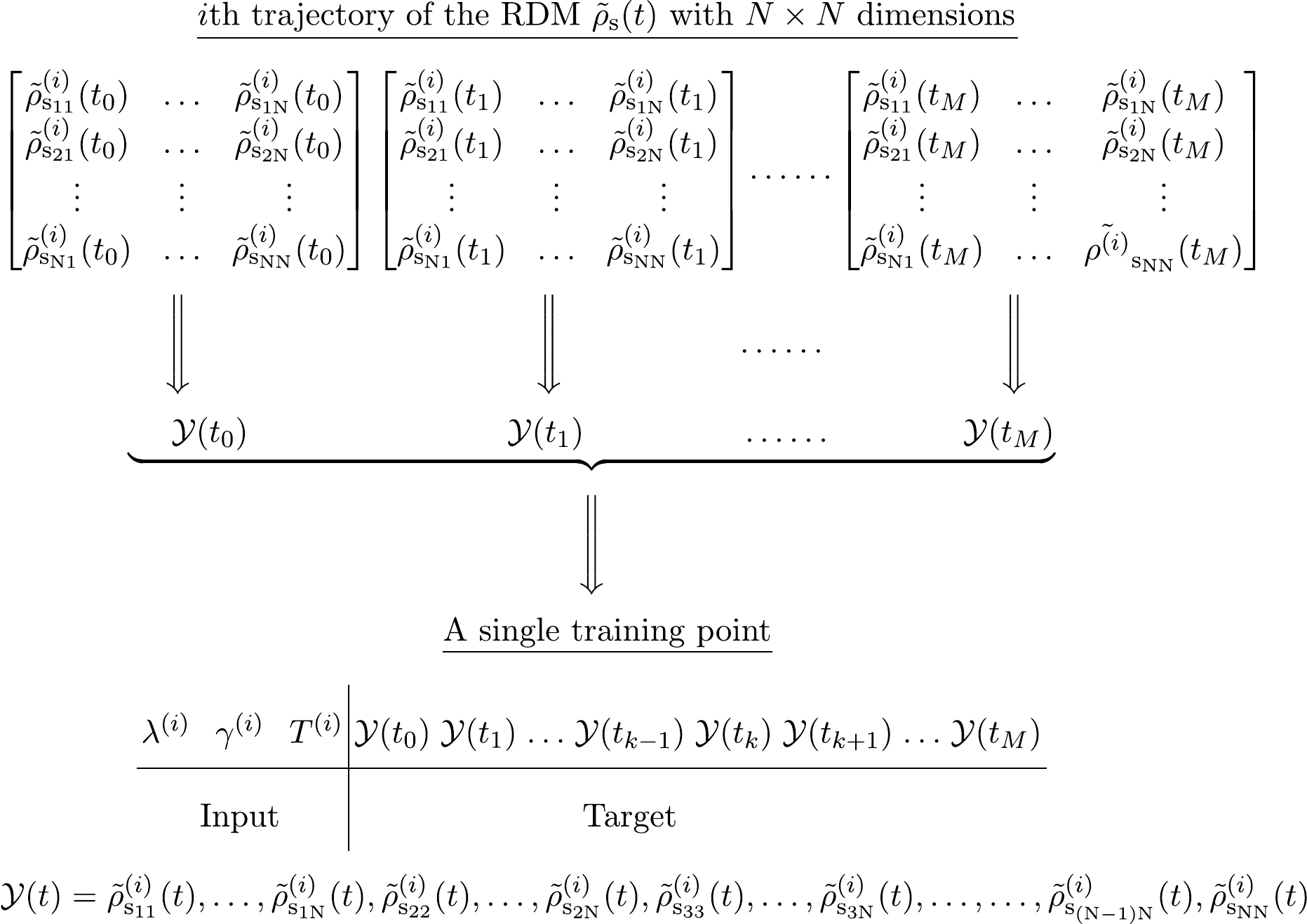}
    \caption{Transformation of the $i$th trajectory of system's RDM $\tilde{\rho}^{(i)}_{\rm s}(t)$ in the OSTL approach. The $\tilde{\rho}^{(i)}_{\rm s}(t)$ at each time step transforms into a vector $\mathbf{\mathcal{Y}}(t)$ with dimension $M$ =  number of sites + (2 $\times$ number of the upper off-diagonal terms). As in RMD $\tilde{\rho}_{\rm s_{nm}}(t) = \tilde{\rho}_{\rm s_{mn}}^*(t) \, (n \neq m)$, only the upper off-diagonal terms are learned. In addition, the real and imaginary parts of each off-diagonal term are separated. Simulation parameters $\lambda^{(i)}$, $\gamma^{(i)}$ and $T^{(i)}$ are the reorganization energy, characteristic frequency, and temperature of the $i$th trajectory in their respective order.}
    \label{fig:ostl}
\end{figure}

\paragraph{OSTL approach.}
Similar to the AIQD approach, in the OSTL approach\cite{ullah2022one} dynamics property (system's state or RDM) can be predicted without step-wise dynamics propagation. However, in contrast to the AIQD, OSTL predicts entire trajectory in one shot for a discretized set of time steps. OSTL also includes the simulation parameters (i.e., $\lambda, \gamma, T$) as an input, while time is not included. As shown in Fig.~\ref{fig:ostl}, in the OSTL approach, each trajectory of system's RDM transforms into a single training point thus significantly reducing the cost of training. The full-time dynamics is predicted using the multi-output feature which obviates invoking the entire ML structure for each time step, thus leading to a significant speed up in dynamics prediction too.       

\subsection{Machine learning models}
\subsubsection{Kernel ridge regression}
\label{subsubsec:krr}
In kernel ridge regression (KRR), for a given input vector $\mathbf{r}$, a function $f(\mathbf{r})$ is approximated with the following expansion~\cite{stulp2015many,dral2019mlatom} 
\begin{equation} \label{eq:krr}
    f(\mathbf{r}) = \sum_{i}^{N_{tr}}\eta_i \mathrm{K}(\mathbf{r}, \mathbf{r}_i), 
\end{equation}
where $N_{tr}$ is the number of training points, $\mathbf{\eta}=\{\eta_i\}$ is a vector of regression coefficients and $\mathrm{K}(\mathbf{r}, \mathbf{r}_i)$ is a kernel function measuring the similarity between two vectors $\mathbf{r}$ and $\mathbf{r}_i$. The very common kernel is the Gaussian kernel\cite{dral2019mlatom,dral2020quantum}
\begin{equation}\label{eq:gaussian}
    \mathrm{K}\left(\mathbf{r}, \mathbf{r}_{i}\right)=\exp \left(-\frac{\left\|\mathbf{r}-\mathbf{r}_{i}\right\|_{2}^{2}}{2 \sigma^{2}}\right) \, ,
\end{equation}
where $\sigma$ is a hyperparameter defining the length scale. It is worth emphasising that many other kernel functions $\mathrm{K}(\mathbf{r}, \mathbf{r}_i)$ such as Mat\'ern and exponential kernels\cite{rasmussen2003gaussian, gneiting2010matern} can also be used, however, based on our previous studies,\cite{ullah2021speeding,rodriguez2022comparative} these kernels do not outperform the Gaussian kernel, thus in \textsc{MLQD}, we only use the Gaussian kernel.

To find the regression coefficients $\mathbf{\eta}$ in Eq.~\eqref{eq:krr}, \textsc{MLQD} uses \textsc{MLatom} package\cite{dral2019mlatom,dral2021mlatom,MLatomdev} in the backend and solves the following equation
\begin{equation} \label{eq:coeff}
\left(\mathbf{K} + \lambda \mathbf{I} \right)\mathbf{\eta} = \mathbf{y},
\end{equation}
where $\mathbf{K}$ is the kernel matrix, $\mathbf{I}$ is the identity matrix, $\mathbf{y}$ is the vector of target values, and $\lambda$ represents a non-negative regularization hyperparameter.

\subsubsection{Neural networks}
\label{subsubsec:nn}
\begin{figure}
    \centering
    \includegraphics[width=0.85\textwidth]{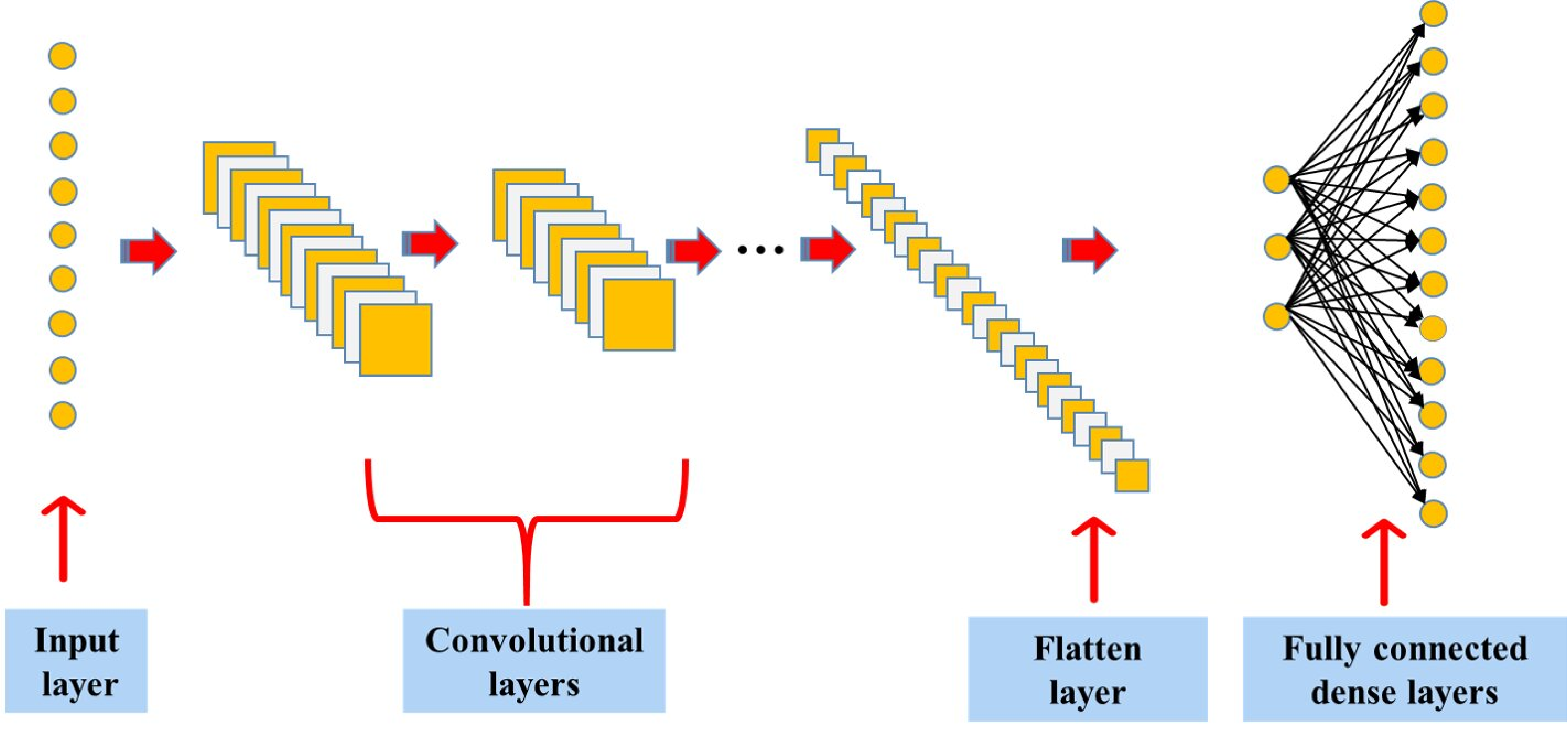}
    \caption{A flowchart of one dimensional CNN model in \textsc{MLQD} which consists an input layer, convolutional layers, a maximum pooling layer (not shown), a flatten layer, dense layers and an output layer.}
    \label{fig:cnn}
\end{figure}

Carefully-constructed neural networks (NN) models, consisting of neurons organized in layers, can be considered as universal approximators of any continuous function.~\cite{hornik89,cybenko89,leshno93} There is a large group of NN models that can be used to learn and predict quantum dynamics,\cite{rodriguez2022comparative} however here we will restrict ourselves to the one dimensional (1D) convolutional neural network (CNN) model as it is the only NN model implemented in the \textsc{MLQD} package. As shown in Fig.~\ref{fig:cnn}, the CNN model in \textsc{MLQD} package comprises of an input layer (expects input in 3D), two to three convolutional layers (in and out in 3D), a maximum pooling layer (outputs in 3D), a flatten layer (outputs in 1D), two to three 1D dense layers (in and out in 1D) and a 1D output layer. In our CNN model, convolutional layers extract the features such as time dependence, the maximum pooling layer decreases the size of the feature map, the flatten layer transforms the output from the maximum pooling layer into one-dimensional vector and then we have the dense layers which are the common feed-forward neural networks. Since the convolutional layers anticipate inputs in a 3D format, it becomes necessary to reshape our training data, which exists in two dimensions (number of training instances, temporal length of each instance), before it can be fed into the network.      

\subsection{Optimization of hyperparameters}
In the \textsc{MLQD} package, we provide a set of default values for hyperparameters, however, the hyperparameter optimization is also possible. In the case of KRR approach, \textsc{MLQD} uses \textsc{MLatom}'s grid search functionality and optimizes its both hyperparameters $\sigma$ and $\lambda$. In the case of AIQD and OSTL approaches, \textsc{MLQD} uses the \textsc{hyperopt} library\cite{Bergstra2015HyperoptAP} for the optimization of CNN structure. \textsc{hyperopt} uses Bayesian optimization with the parallel infrastructure for a fast search of best hyperparameters in a defined multidimensional space. We optimize the number of filters, kernels size, the number of neurons, learning rate and the number of batches in a predefined multidimensional space. The number of hidden convolutional layers and hidden dense layers are optimized between two numbers, i.e., \{2, 3\}.

\subsection{Plotting}
For a better understanding of any approach, visualization of results is necessary and in many cases such as cloud computing, auto-plotting provides complete mouse-click computing. We incorporate this functionality in \textsc{MLQD}, where the predicted dynamics is plotted against the reference trajectory providing a clear visualization of the predicted dynamics.        

\section{Applications} \label{sec:applications}
In this section, we present two case studies to highlight the applications of the \textsc{MLQD} package. We consider the two widely studied systems, namely the spin-boson model and the 8-site Fenna--Matthews--Olson (FMO) complex.

Before that, we briefly overview the general theory behind open quantum systems. A quantum system coupled to its outside environment (bath) is regarded as an open quantum system with the dynamics governed by the following Hamiltonian 
\begin{equation} \label{eq:totH}
    \mathbf{H} = \mathbf{H}_{\rm s} + \mathbf{H}_{\rm b} + \mathbf{H}_{\rm sb},
\end{equation}
where $\mathbf{H}_{\rm s}$ and $\mathbf{H}_{\rm b}$ represent the Hamiltonian for the system and the outside environment (bath), respectively. The last term $\mathbf{H}_{\rm sb}$ incorporates the interaction between the system and the environment. To propagate quantum dynamics, Liouville--von Neumann equation can be employed
\begin{equation}
    \dot{\mathbf{\rho}}(t) = -i[\mathbf{H}, \mathbf{\rho}(t)], 
\end{equation}
where $\mathbf{\rho}(t)$ is the density matrix at time $t$ and $\hbar$ is set to 1. In system-bath approaches,\cite{tanimura1989time, yan2004hierarchical,xu2015dissipaton,shao2004decoupling} calculations are usually simplified by considering system and environment uncorrelated at $t=0$, i.e., $\mathbf{\rho}(0) = \mathbf{\rho}_{\rm s}\mathbf{\rho}_{\rm b}$ where $\mathbf{\rho}_{\rm s}$ is the density matrix of the system and $\mathbf{\rho}_{\rm b}$ denotes density matrix of the environment. As we are interested only in the system, we can take a partial trace over environment degrees of freedom
\begin{equation}
    \tilde{\mathbf{\rho}}_{\rm s} = \mathbf{Tr}_{\rm b} \left[ \mathbf{U}(t,0) \mathbf{\rho}(0) \mathbf{U}^\dagger(t,0)\right],
\end{equation}
where $\tilde{\mathbf{\rho}}_{\rm s}$ is the density matrix of the reduced system (reduced density matrix (RDM)) and $\mathbf{U}(t,0)\left(\mathbf{U}^\dagger(t,0)\right)$ is the propagation operator forward (backward) in time and $\mathbf{Tr}_{\rm b}$ is the partial trace over environment degrees of freedom. In reality, the term \textit {open quantum system} is applicable to most of the systems, however, because of the curse of dimensionality, not all of them are easy to be theoretically handled. In the following, we present a brief theory of two broadly studied pedagogical systems, the two-states spin-boson model and the FMO complex. 

\subsection{Case study 1: Relaxation dynamics of spin-boson model}
As a first case study, we consider the relaxation dynamics of excited state $|e\rangle$ in spin-boson model. The spin-boson model is a two-state system coupled with an environment of an infinite number of non-interacting harmonic oscillators. The Hamiltonian of the composite system (two-states system + environment) is expressed as 
\begin{equation}\label{eq:sb}
    \mathbf{H} = \epsilon \left(|e\rangle\langle e| - |g\rangle\langle g| \right) + \Delta \left(|e\rangle\langle g| + |g\rangle\langle e| \right) + \sum_{k=1} \omega_k \mathbf{b}^\dagger_k \mathbf{b}_k + \left(|e\rangle\langle e| - |g\rangle\langle g| \right) \sum_{k=1} c_k \left(\mathbf{b}^\dagger + \mathbf{b}_k\right),
\end{equation}
where $|e\rangle$ and $|g\rangle$ denote the two states of the system, $\epsilon$ is the half of the energy difference between the two states and $\Delta$ is the tunneling splitting. The $\mathbf{b}^\dagger_k (\mathbf{b}_k)$ denotes the creation (annihilation) operator in the environment Hilbert space and $\omega_k$ is the frequency corresponding to $k$ mode. The last term in Eq.~\eqref{eq:sb} incorporates the interaction between the system and environment with $c_k$ as the coupling strength between the system's operator and $k$ environment mode. The effects of the environment on system dynamics are described by the spectral density of the environment 
\begin{equation}
    J(\omega) =\sum_k\alpha_k \delta(\omega-\omega_k),
\end{equation}
where $\alpha_k = \frac{\pi}{2} \frac{c_k^2}{m_k \omega_k}$. Here we adopt the Ohmic spectral density function with the Drude--Lorentz cut-off\cite{caldeira1983path}
\begin{equation}\label{eq:drude}
    J(\omega) = 2 \lambda \frac{\gamma \omega}{\omega^2 +  \gamma^2},
\end{equation}
where $\lambda$ is the reorganization energy and $\gamma$ is the characteristic frequency or the inverse of environment relaxation time, i.e., $\gamma = 1/\tau$.   

For our example, we use the spin-boson data set from our recently published QD3SET-1 database.\cite{ullah2023qd3set} The mentioned data set consists of 1000 trajectories generated for each possible combination of the following parameters; $\tilde{\epsilon}=\epsilon/\Delta=\{0, 1\}$, $\tilde{\lambda}=\lambda/\Delta=\{0.1, 0.2, 0.3, 0.4, 0.5, 0.6,0.7, 0.8, 0.9, 1.0\}$, 
$\tilde{\gamma}=\gamma/\Delta =\{1, 2, 3, 4, 5, 6, 7, 8, 9, 10\}$, and $\tilde{\beta}=\beta \Delta=\{0.1, 0.25, 0.5, 0.75, 1\}$,
where the tunneling matrix element $\Delta$ is set as an energy unit. Data is generated with the HEOM method implemented in the \textsc{QuTiP} software package\cite{johansson2012qutip} and each trajectory is propagated up to $t\Delta = 20$ with the time step $dt\Delta = 0.05$. 

In this case study, we use all three available approaches (i.e., KRR, AIQD and OSTL) to predict the relaxation dynamics in the two possible cases, namely the symmetric case $\tilde{\epsilon} = 0$ and the asymmetric case $\tilde{\epsilon} = 1.0$. Before training, we divide the spin-boson data set into the training set (400 trajectories for each case) and the test set (100 trajectories for each case). The division is based on farthest-point sampling which selects the most distant points in a three-dimensional Euclidean space\cite{ullah2022predicting} $(\tilde{\lambda}, \tilde{\gamma}, \tilde{\beta})$, thus efficiently covering the parameter space in comparison to random sampling.\cite{dral2019mlatom} Keeping in mind the high computational cost of KRR, we sample the data for training with comparatively larger time step $dt\Delta = 0.1$.

\begin{figure}
    \centering
    \includegraphics[width=\textwidth]{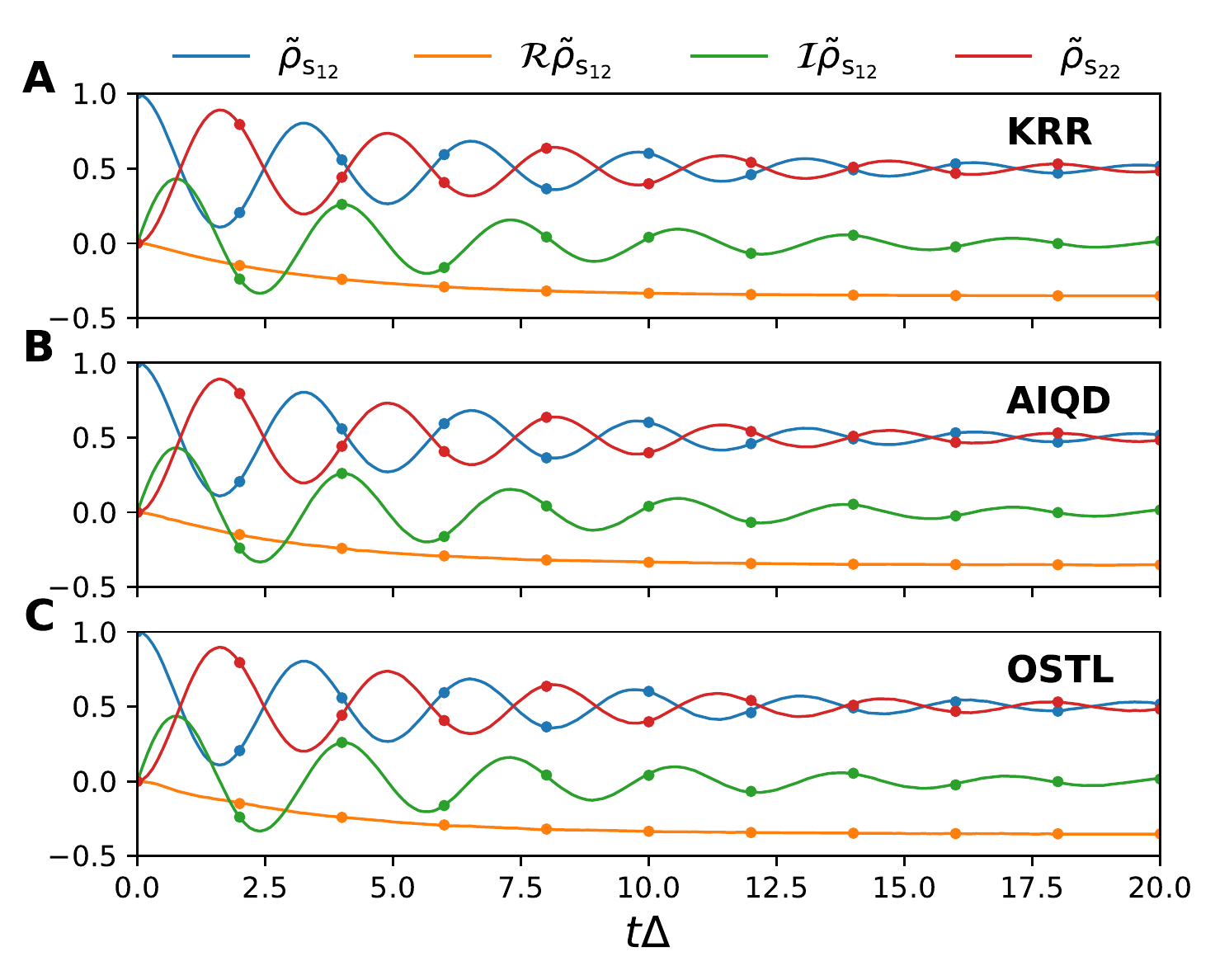}
     \caption{The time evolution of the diagonal (population) and off-diagonal terms (coherence) of the reduced density matrix (RDM) $\tilde{\rho}_{\rm s}$ predicted with (A) the recursive (KRR) and non-recursive (B) AIQD and (C) OSTL approaches. The calligraphic $R$ and $I$ denote the real and imaginary parts of the off-diagonal terms, respectively. Results are shown for the symmetric spin-boson model ($\Tilde{\epsilon} = 0.0$) with the following set of unseen parameters: $\tilde{\gamma} = 10.0$, $\tilde{\lambda}=0.3$, $\tilde{\beta}=1.0$. The predicted results are compared to the reference HEOM method (dots).}
     \label{fig:sb_sym}
\end{figure}

For KRR, a short-time trajectory of $t_m\Delta = 4.0$ is used as an input and following the algorithm described in Fig.~\ref{fig:krr},\cite{ullah2021speeding} we transform the  trajectories beyond $t_m\Delta = 4.0$ into target values.
We train separate KRR models for each diagonal element of the RDM. As $\tilde{\mathbf{\rho}}_{\rm s_{12}} = \tilde{\mathbf{\rho}}_{\rm s_{21}}^*$, we learn only the upper off-diagonal term where we train a separate KRR model for real and imaginary parts. After training, we provide a reference short-time trajectory of time-length $t_m\Delta = 4.0$ to initiate recursive propagation with the trained KRR model (the recursive propagation is beyond this short-time trajectory dynamics). In Fig.~\ref{fig:sb_sym}(A), we show population and coherence for symmetric case $\tilde{\epsilon} = 0$ while results for asymmetric case $\tilde{\epsilon} = 1$ are shown in Fig.~{\color{blue} S1}(A).  

In the case of AIQD and OSTL approaches, we prepare training data following Figs.~\ref{fig:aiqd} and \ref{fig:ostl} in their respective order. In both approaches, we use $\tilde{\gamma}_{\rm max} = 10.0$ for $\tilde{\gamma}$ and 1.0 for the remaining simulation parameters as normalization factors. To normalize the dimension of time, in AIQD we use a set of 10 logistic functions, i.e., $f_j(t) = a/\left(1 + b\exp\left(-(t+c_j)/d\right)\right)$ where $j = 0,1,2, \dots, 9$. We set $a = 1.0$, $b = 15.0$, $d = 1.0$ and $c = 5j - 1$. In List.~\ref{lst:mlqd-ostl-1}, we show an example of \textsc{MLQD} input for creating (training) a CNN model following the OSTL approach. After training, by providing the trained AIQD model, values of the simulation parameters and time, \textsc{MLQD} predicts the corresponding RDM $\tilde{\rho}_{\rm s}$. In the OSTL approach, the RDM $\tilde{\rho}_{\rm s}$ is predicted for the whole time range, i.e., $t_m\Delta = 0, \dots, \dots t_m\Delta = 20$. In Figs.~\ref{fig:sb_sym}(B) and (C), we show the time evolution of RDM's diagonal (population) and off-diagonal terms (coherence) for a set of test parameters. Results for the asymmetric case $\tilde{\epsilon} = 1$ are given in Figs.~{\color{blue} S1}(B) and (C).   

\begin{lstlisting}[caption={An illustration of an \textsc{MLQD} input used to train a CNN model using the OSTL approach. In this configuration, the parameter \lstinline{n_states} specifies the number of states. The \lstinline{QDmodel} feature instructs \textsc{MLQD} to generate a QD model through training, while \lstinline{QDmodelType} defines the approach type. By activating the \lstinline{prepInput} option, trajectory preprocessing is initiated following the procedure outlined in Fig.~\ref{fig:ostl}. The processed training data is saved with the file names 'x\_data' and 'y\_data' controlled by the \lstinline{XfileIn} and \lstinline{YfileIn} parameters, respectively. Normalization factors such as \lstinline{energyNorm}, \lstinline{DeltaNorm}, \lstinline{gammaNorm}, \lstinline{lambNorm} and \lstinline{tempNorm} are employed to scale $\epsilon$, $\Delta$, $\gamma$, $\lambda$ and $T$ ($\beta$ in the case of SB model) respectively. The \lstinline{systemType} parameter designates the required system type for input preparation. Optimization of hyperparameters can be enabled or disabled through the \lstinline{hyperParam} parameter set to either 'True' or 'False'. The \lstinline{patience} value determines the early stopping patience, while \lstinline{OptEpochs} and \lstinline{TrEpochs} define the optimization and training epochs, respectively. The parameter \lstinline{max_evals} sets the limit for optimization evaluations. The \lstinline{dataPath} parameter points to the directory containing the training trajectories, and \lstinline{QDmodelOut} ensures that the trained model is saved as 'OSTL\_CNN\_SB\_model'.}, label={lst:mlqd-ostl-1},  captionpos=b]

from mlqd.evolution import quant_dyn
param={ 
        'n_states': 2,         
        'QDmodel': 'createQDmodel',
        'QDmodelType': 'OSTL', 
        'prepInput' : 'True',  
        'XfileIn': 'x_data',
        'YfileIn': 'y_data',
        'energyNorm': 1.0,
        'DeltaNorm': 1.0,
        'gammaNorm': 10.0,
        'lambNorm': 1.0,
        'tempNorm': 1.0,
        'systemType': 'SB',
        'hyperParam': 'True',
        'patience': 100,
        'OptEpochs': 100,
        'TrEpochs': 1000,
        'max_evals': 50,
        'dataPath': '/path/to/training_trajectories',
        'QDmodelOut': 'OSTL_CNN_SB_model'
        }
quant_dyn(**param)
\end{lstlisting}

\subsection{Case study 2: Excitation energy transfer in FMO complex with AIQD approach}
In our second case study, we consider excitation energy transfer in the FMO complex which is described by the Frenkel exciton model with the following Hamiltonian\cite{ishizaki2009unified}   
\begin{align} \label{eq:fmo_hamil}
    \mathbf{\rm H} = &   \sum_{n=1}^{N} |n\rangle\epsilon_n \langle n| + \sum_{n,m=1, n\neq m}^{N} |n \rangle J_{nm} \langle m| + \sum_{n=1}^{N} \sum_{k=1} \left(\frac{1}{2} \mathbf{\rm P}_{k, n}^{2} + \frac{1}{2} \omega_{k, n}^{2} \mathbf{\rm Q}_{k, n}^{2}\right) \mathbf{I}\nl 
    & - \sum_{n=1}^{N} \sum_{k=1} | n \rangle c_{k,n} \mathbf{\rm Q}_{k, n} \langle n| + \sum_{n=1}^{N} | n \rangle \lambda_{n} \langle n | \, , 
\end{align}
where $N$ is the number of sites (bacteriochlorophyll molecules), $\epsilon_n$ is the energy of the $n$th site and $J_{nm}$ denotes the inter-site coupling between sites $n$ and $m$. The third term in Eq.~\eqref{eq:fmo_hamil} describes the environmental part with $\mathbf{\rm P}_{k, n}$ as conjugate momentum, $\mathbf{\rm Q}_{k, n}$ as coordinate and $\omega_{k, n}$ as a frequency of the corresponding environment mode $k$. $\mathbf{I}$ is $N × N$ identity matrix that has been multiplied to ensure consistency in dimensions. $\lambda_{n}$ is the reorganization energy associated with site $n$ and the strength of the coupling between environment mode $k$ and site $n$ is represented by $c_{k,n}$. In the case of FMO complex, we assume that all sites have the same spectral density as described by Eq.~\eqref{eq:drude}. 

In our example, we take the 8-site FMO complex where three sites (1, 6 and 8) have an equal probability of getting initially excited while the reaction center is in the vicinity of sites 3 and 4.
For training, we use the FMO-IV data set from the QD3SET-1 database\cite{ullah2023qd3set} generated for the following Hamiltonian\cite{schmidt2011eighth, olbrich2011atomistic}
\begin{equation}
    H_s = \begin{pmatrix}
310 & -80.3 & 3.5 & -4.0 & 4.5 & -10.2 & -4.9 & 21.0\\
-80.3 & 230 & 23.5 & 6.7 & 0.5 & 7.5 & 1.5 & 3.3\\
3.5 & 23.5 & 0 & -49.8 & -1.5 & -6.5 & 1.2 & 0.7\\
-4.0 & 6.7 & -49.8 & 180 & 63.4 & -13.3 & -42.2 & -1.2\\
4.5 & 0.5 & -1.5 & 63.4 & 450 & 55.8 & 4.7 & 2.8\\
-10.2 & 7.5 & -6.5 & -13.3 & 55.8 & 320 & 33.0 & -7.3\\
-4.9 & 1.5 & 1.2 & -42.2 & 4.7 & 33.0 & 270 &-8.7\\
21.0 & 3.3 & 0.7 & -1.2 & 2.8 & -7.3 & -8.7 & 505
\end{pmatrix}, \label{eq:fmo2-1}
\end{equation}
with the diagonal offset of 12195 cm$^{-1}$. In the considered data set, exciton dynamics is propagated for the most distant 500 combinations of the following parameters: $\lambda$ = $\{$10, 40, 70, \ldots, 520$\}$ cm$^{-1}$,
$\gamma$ = $\{$25, 50, 75, \ldots, 500$\}$ cm$^{-1}$,
and T = $\{$30, 50, 70, \ldots, 510$\}$ K. The chosen 500 trajectories are propagated for each possible case of initial excitation (i.e., on sites 1, 6 and 8) with time length $t = 50$~ps and time step $dt= 5$~fs. Calculations are performed with the local thermalizing Lindblad master equation (LTLME) approach,\cite{mohseni2008environment, joseph2020thesis} implemented in the \textsc{quantum\_HEOM} package.\cite{joseph2019quant} In order to make it compatible with the Hamiltonians with larger dimensions, the package is modified locally. 

\begin{figure}
    \centering
    \includegraphics[width=\textwidth]{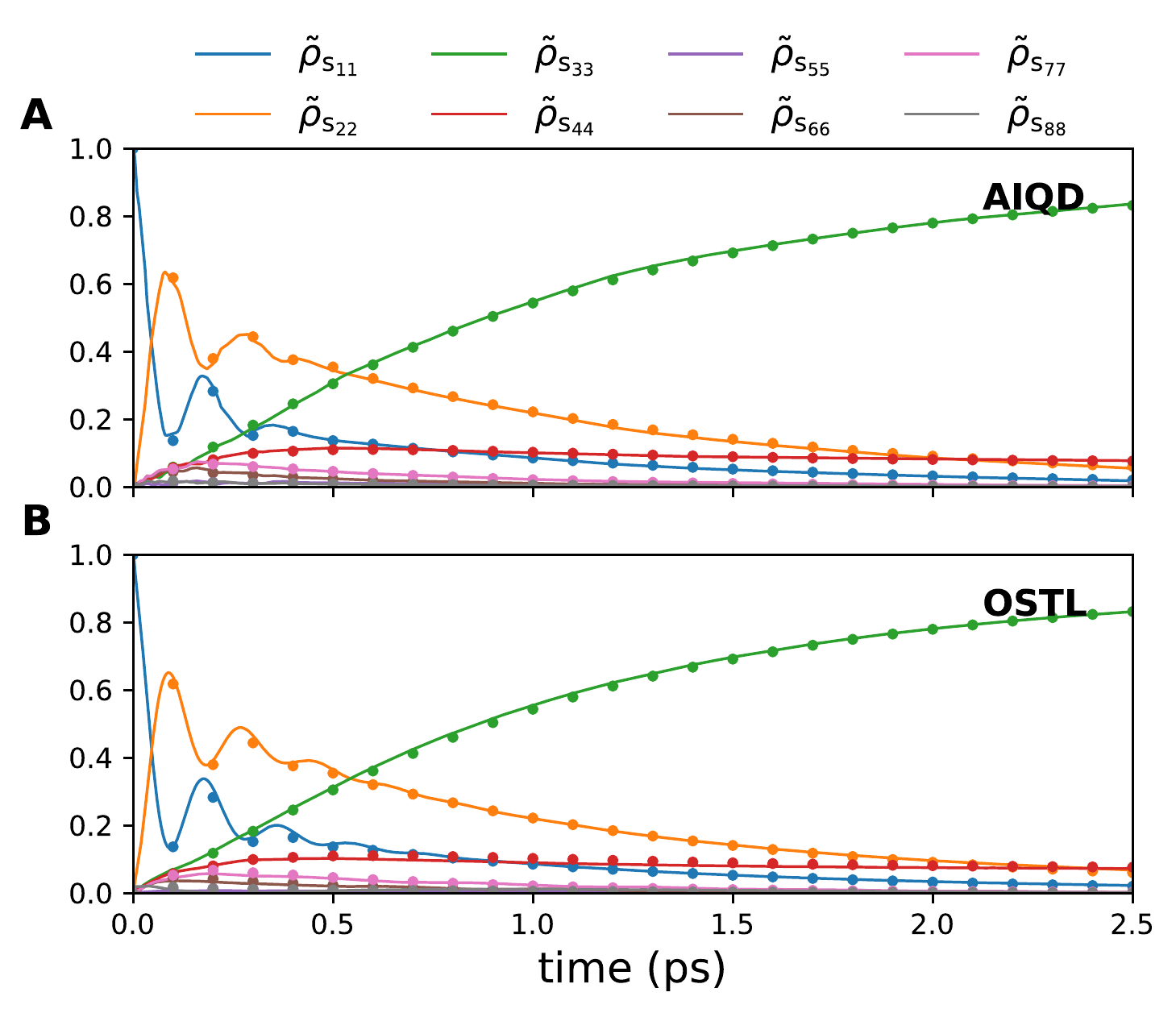}
    \caption{The time evolution of the diagonal elements of RDM $\tilde{\rho_{\rm s}}$ with (A) AIQD and (B) OSTL approaches. The initial excitation is on site 1 and other parameters are $\gamma = 125$, $\lambda =70$, $T=30$. The results are compared to the reference LTLME method (dots). In our calculations, $\gamma$ and $\lambda$ are considered in the units of cm$^{-1}$, while $T$ is in the units of K. The time evolution of the prominent off-diagonal terms can be found in Fig.~\ref{fig:site1_coherence}.}
    \label{fig:site1_pop}
\end{figure}

\begin{figure}
    \centering
    \includegraphics[width=\textwidth]{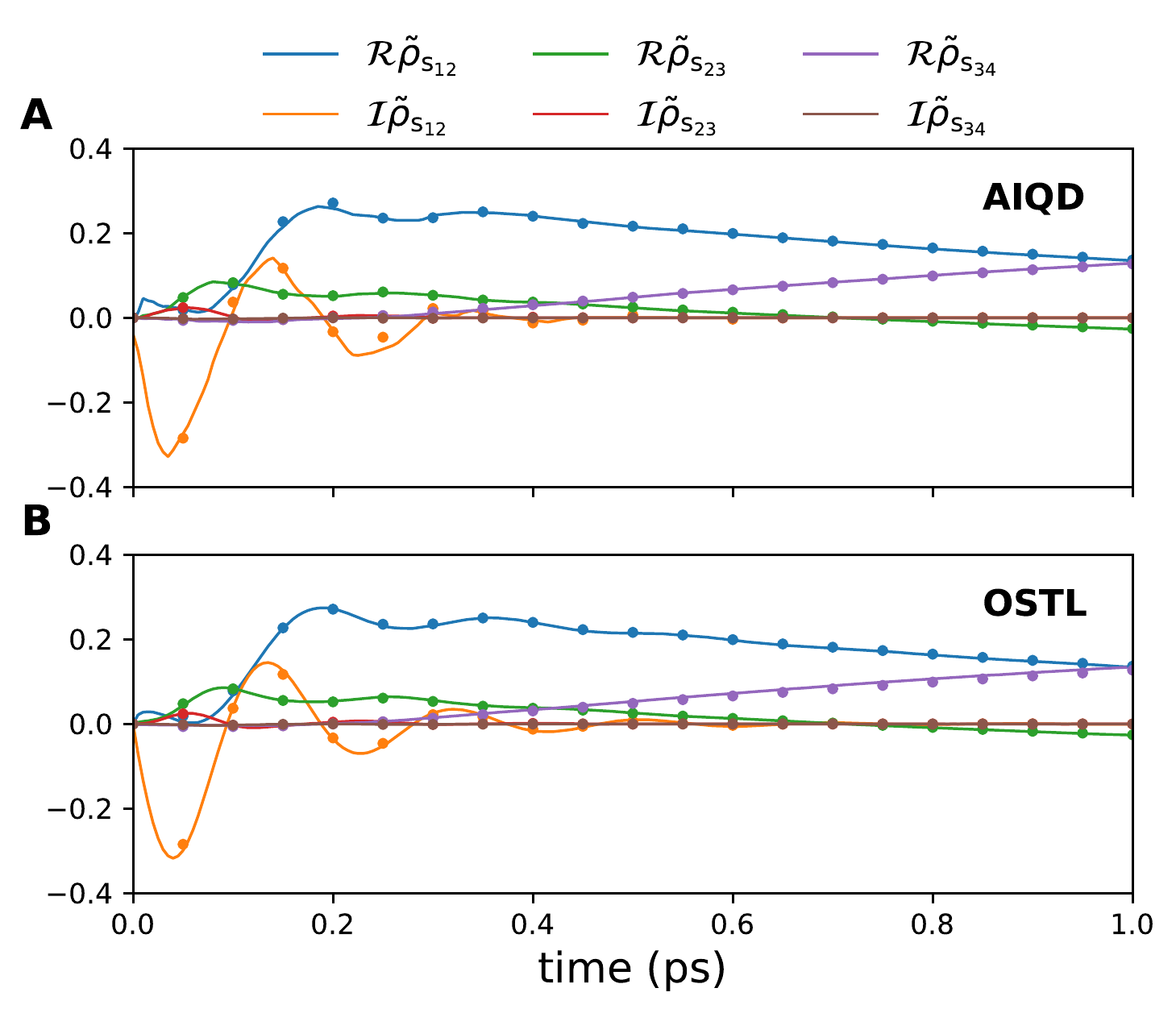}
    \caption{The time evolution of the dominant off-diagonal terms (i.e., $\tilde{\rho}_{\rm s_{mn}}, m\neq n$) with the initial excitation on site-1, predicted with the (A) AIQD and (B) OSTL approaches. The calligraphic $R$ and $I$ represent the real and imaginary parts, respectively. The results are compared to the reference LTLME method (dots). The time evolution of diagonal terms and the corresponding simulation parameters are given in Fig.~\ref{fig:site1_pop}.}
    \label{fig:site1_coherence}
\end{figure}

We choose our training trajectories (400 for each case of initial excitation) and test trajectories (100 for each case of initial excitation) based on farthest-point sampling. In this case study, we consider only the AIQD and OSTL approaches as training the KRR model (in the current implementation) is not feasible because of its high computational cost. In both approaches (AIQD and OSTL), we use $\lambda_{\rm max} = 520$, $\gamma_{\rm max} = 500$ and $T_{\rm max} = 510$ as normalizing factors for the corresponding simulation parameters (i.e., $\lambda, \gamma$ and $T$). In AIQD, we use the same number of logistic functions and the same constants as was adopted for the spin-boson case. After training, we pass a set of unseen simulation parameters and \textsc{MLQD} predicts the corresponding dynamics using the trained CNN models. In the described order, Figs~\ref{fig:site1_pop} and \ref{fig:site1_coherence} show the excitation energy transfer (diagonal terms of RDM) and the time evolution of coherent terms (off-diagonal terms of RDM) for a test trajectory. The presented results are with the initial excitation on site 1 and results for the initial excitation on sites 6 and 8 are given in Figs.~{\color{blue} S2}-{\color{blue}5}. In List.~\ref{lst:mlqd-ostl-2}, we show an example of the \textsc{MLQD} input for predicting exciton dynamics in the FMO complex using the OSTL approach.   
\begin{lstlisting}[caption={A demonstration of \textsc{MLQD} input utilized to forecast exciton dynamics within the FMO complex using the OSTL approach. In this instance, \lstinline{initState} designates the site holding the initial excitation, while \lstinline{n_states} determines the number of states (sites) involved. The parameter \lstinline{time} specifies the time for propagating dynamics, with the time step provided as \lstinline{time_step}. By incorporating the \lstinline{QDmodel} feature, \textsc{MLQD} is directed to utilize a pre-existing QD model. The \lstinline{QDmodelType} parameter outlines the chosen approach for dynamics propagation. Parameters \lstinline{gamma}, \lstinline{lamb} and \lstinline{temp} are employed for defining the simulation parameters ($\gamma$, $\lambda$, $T$) for which dynamic predictions are sought. \lstinline{systemType} is used to specify the system type, which isn't limited to SB or FMO. Nevertheless, \textsc{MLQD} includes preconfigured parameters for SB and FMO. Additionally, \lstinline{QDmodelIn} provides a pre-trained QD model to be employed in the prediction of dynamics.}, label={lst:mlqd-ostl-2},  captionpos=b]

from mlqd.evolution import quant_dyn
param={
        'initState': 8,
        'n_states': 8,
        'time': 50,
        'time_step': 5,
        'QDmodel': 'useQDmodel',
        'QDmodelType': 'OSTL',
        'gamma': 200.0,
        'lamb': 130.0,
        'temp': 330.0,
        'systemType': 'FMO',
        'QDmodelIn': 'OSTL_CNN_FMO_model.hdf5'
        }
quant_dyn(**param)
\end{lstlisting}

\section{Conclusions and Outlook}
In this article, we have presented \textsc{MLQD}, an open-source Python package for ML-based quantum dissipative dynamics. The package provides a set of recursive (based on KRR method\cite{ullah2021speeding}) and non-recursive (AIQD\cite{ullah2022predicting} and OSTL\cite{ullah2022one}) ML approaches with the features of training an ML model, using the trained model to predict dynamics, optimization of hyperparameters and visualization of results. The package has been made available on the XACS cloud computing platform with the interface to the \textsc{MLatom} package. 

To highlight the applications of the \textsc{MLQD} package, we demonstrated the features of the \textsc{MLQD} package on examples of the 2-state spin-boson model and 8-site FMO complex, but it is not restricted to them and can be used for any system if the training data is provided. To underscore, machine learning (ML)-based approaches exhibit the capacity to not only interpolate but also extrapolate beyond the confines of the training region to a good extent. Within the context of our study, elaborated in Ref.~\citenum{ullah2021speeding}, we have presented evidence supporting the capability of ML models to extrapolate in the temporal dimension. Furthermore, as expounded upon in the supplementary materials of Ref.~\citenum{ullah2022one}, our investigation of the OSTL approach has effectively showcased successful extrapolation beyond the defined limits of the parameter space. Nevertheless, it is crucial to acknowledge that as extrapolation extends further from the training region, deviations become more pronounced.

To accentuate the advantages of ML methods incorporated in the \textsc{MLQD} package, a comparative study would be of interest and we are currently working on it. In addition, addressing limitations and issues of ML-approaches such as error accumulation and  computational cost of KRR approach, lack of extrapolation in OSTL and finding better ways of time incorporation in AIQD are part of our on-going research. In the future, MLQD will be extended to more realistic systems along with the incorporation of other machine learning-based approaches as well as the traditional quantum dynamics methods such as HEOM and SEOM.    

\section{Supplementary Material}
In Supplementary Material, we provide results for asymmetric spin-boson model as well as for the FMO complex with initial excitations on sites 6 and 8.

\section*{Code Availability Statement}
\textsc{MLQD} package is available on \hyperlink{https://github.com/Arif-PhyChem/MLQD}{https://github.com/Arif-PhyChem/MLQD} along with tutorials in Jupyter Notebooks. It can also be installed as a pip package, i.e., \lstinline{pip install mlqd}. In addition, \textsc{MLQD} is interfaced with the \textsc{MLatom@XACS} package \url{http://mlatom.com} which allows \textsc{MLQD} to be used on the \textsc{XACS} cloud computing platform \url{https://xacs.xmu.edu.cn}. A user manual for \textsc{MLQD} on cloud computing is provided at \url{http://mlatom.com/manual/\#mlqd}.   

\section*{Data Availability Statement}
Data sharing is not applicable to this article as no new data were created or analyzed in this study.

\begin{acknowledgments}
We acknowledge funding by the National Natural Science Foundation of China (No.~22003051 and funding via the Outstanding Youth Scholars (Overseas, 2021) project), the Fundamental Research Funds for the Central Universities (No.~20720210092), and via the Lab project of the State Key Laboratory of Physical Chemistry of Solid Surfaces. This project is supported by Science and Technology Projects of Innovation Laboratory for Sciences and Technologies of Energy Materials of Fujian Province (IKKEM) (No:~RD2022070103). We acknowledge the use of high-performance computing resources provided by Xiamen University for conducting the calculations. We extend our gratitude to Lina Zhang for her assistance in testing the installation and usage of \textsc{MLQD} with the provided examples.  
\end{acknowledgments}

\section*{References}
\bibliography{main}

\end{document}


\begin{suppfigure}
     \centering
     \includegraphics[width=\textwidth]{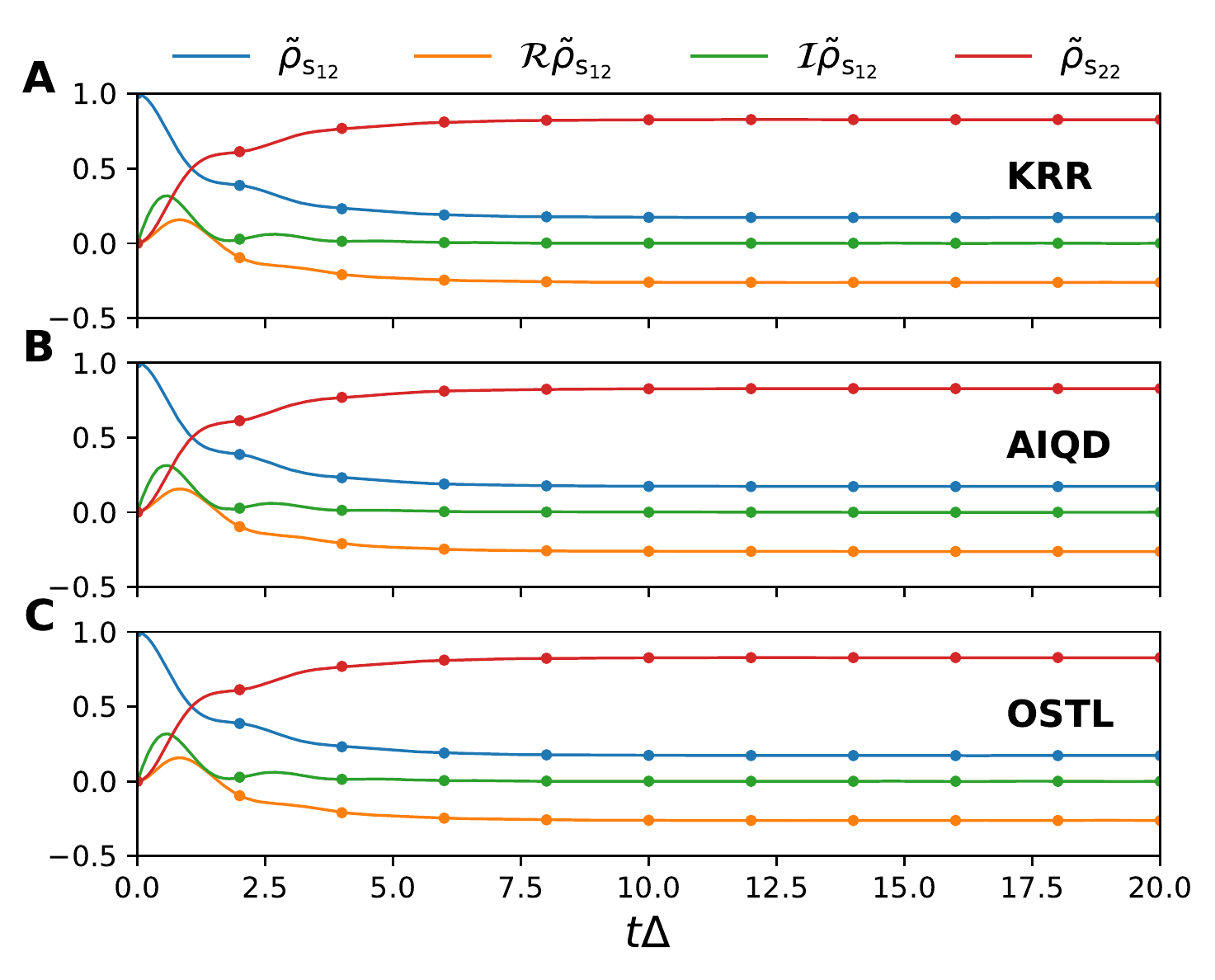}
     \caption{The time evolution of the diagonal (population) and off-diagonal terms (coherence) of the reduced density matrix (RDM) $\tilde{\rho}_{\rm s}$ predicted with both (A) recursive (KRR) and (B)-(C) non-recursive (AIQD and OSTL) approaches. In their respective order, the calligraphic $R$ and $I$ denote the real and imaginary parts of the off-diagonal terms. Results are shown for the asymmetric spin-boson model ($\Tilde{\epsilon} = 1.0$) with the following set of unseen parameters: $\tilde{\gamma} = 3.0$, $\tilde{\lambda}=0.5$, $\tilde{\beta}=1.0$. The predicted results are compared to the reference HEOM method (dots).}
     \label{fig:sb_asym}
 \end{suppfigure}
\begin{suppfigure}
    \centering
    \includegraphics[width=\textwidth]{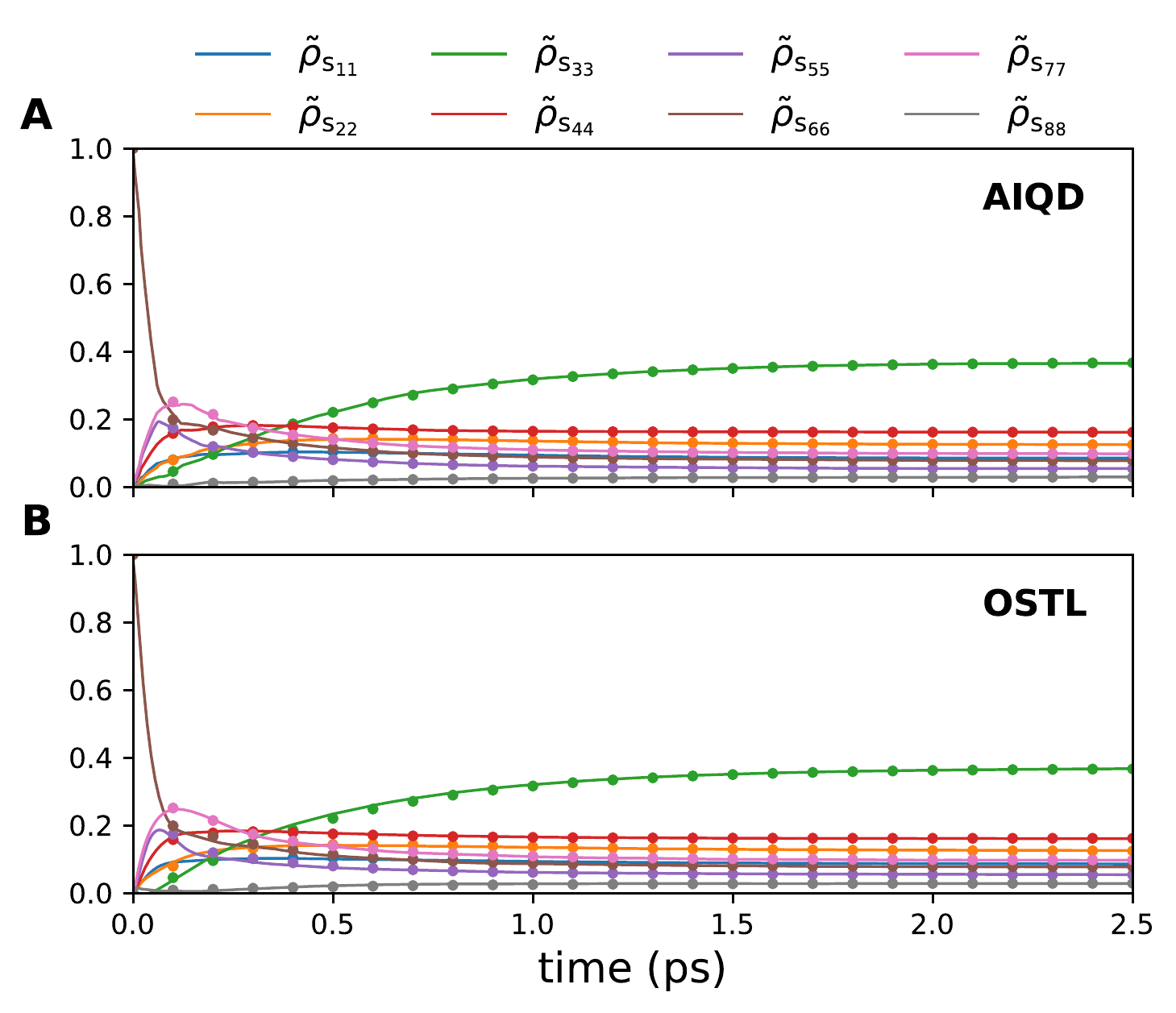}
    \caption{Time evolution of the diagonal elements of RDM $\tilde{\rho_{\rm s}}$ using (A) AIQD and (B) OSTL approaches. The initial excitation is on site 6 and other parameters are $\gamma = 150$~cm$^{-1}$, $\lambda =40$~cm$^{-1}$, $T=290$~K. The results are compared to the reference LTLME method (dots). For the time evolution of the prominent off-diagonal elements, the readers are referred to Fig.~\ref{fig:site6_coherence}.}
    \label{fig:site6_pop}
\end{suppfigure}
 \begin{suppfigure}
    \centering
    \includegraphics[width=\textwidth]{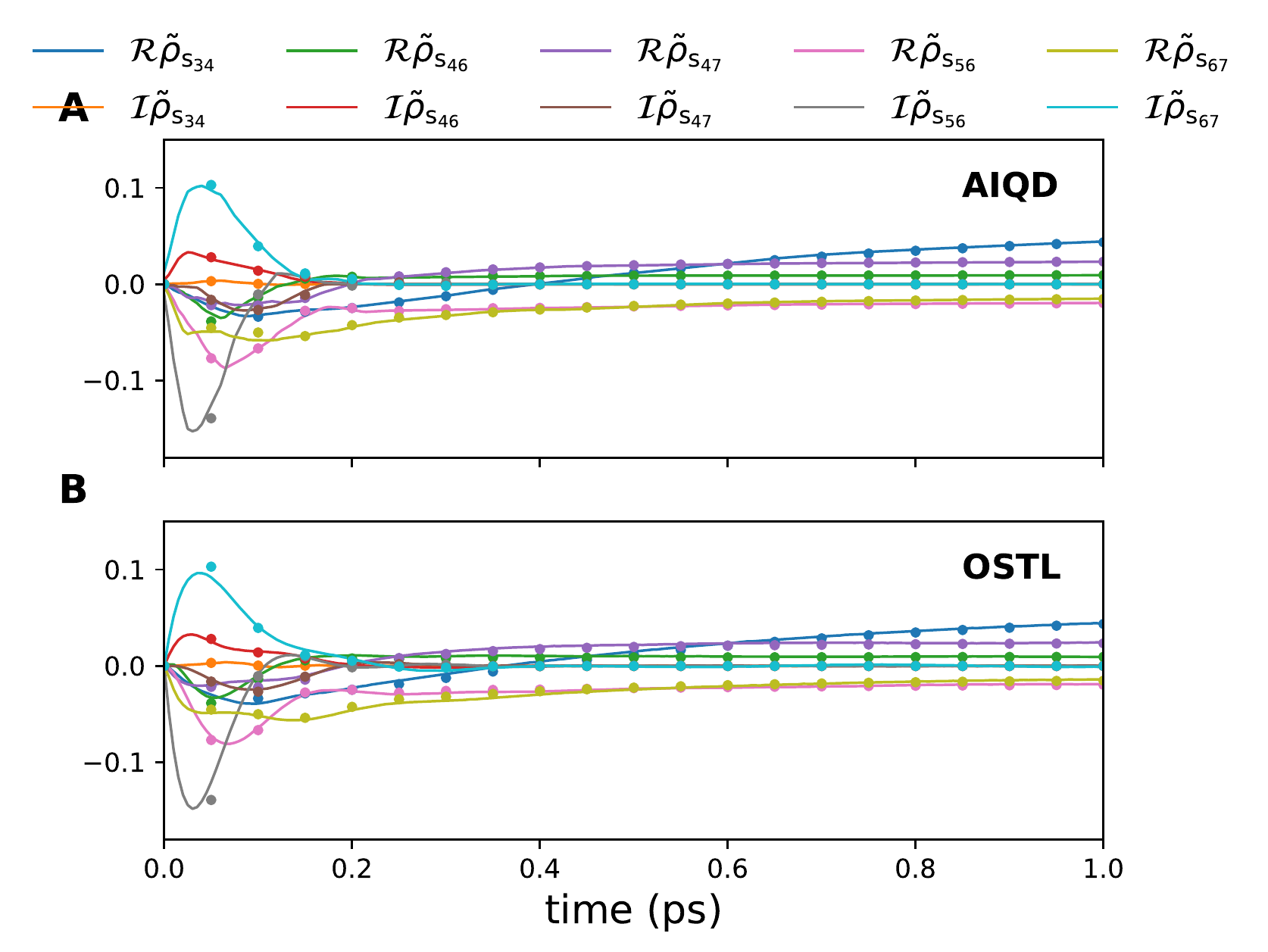}
    \caption{Time evolution of the dominant off-diagonal terms (i.e., $\tilde{\rho}_{\rm s_{mn}}, m\neq n$) using (A) AIQD and (B) OSTL approaches. The initial excitation is considered on site 6. The calligraphic $R$ and $I$ represent the real and imaginary parts, respectively. The results are compared to the reference LTLME method (dots). The time evolution of diagonal terms and the corresponding simulation parameters are given in Fig.~\ref{fig:site6_pop}.}
    \label{fig:site6_coherence}
    \label{fig:}
\end{suppfigure}
\begin{suppfigure}
    \centering
    \includegraphics[width=\textwidth]{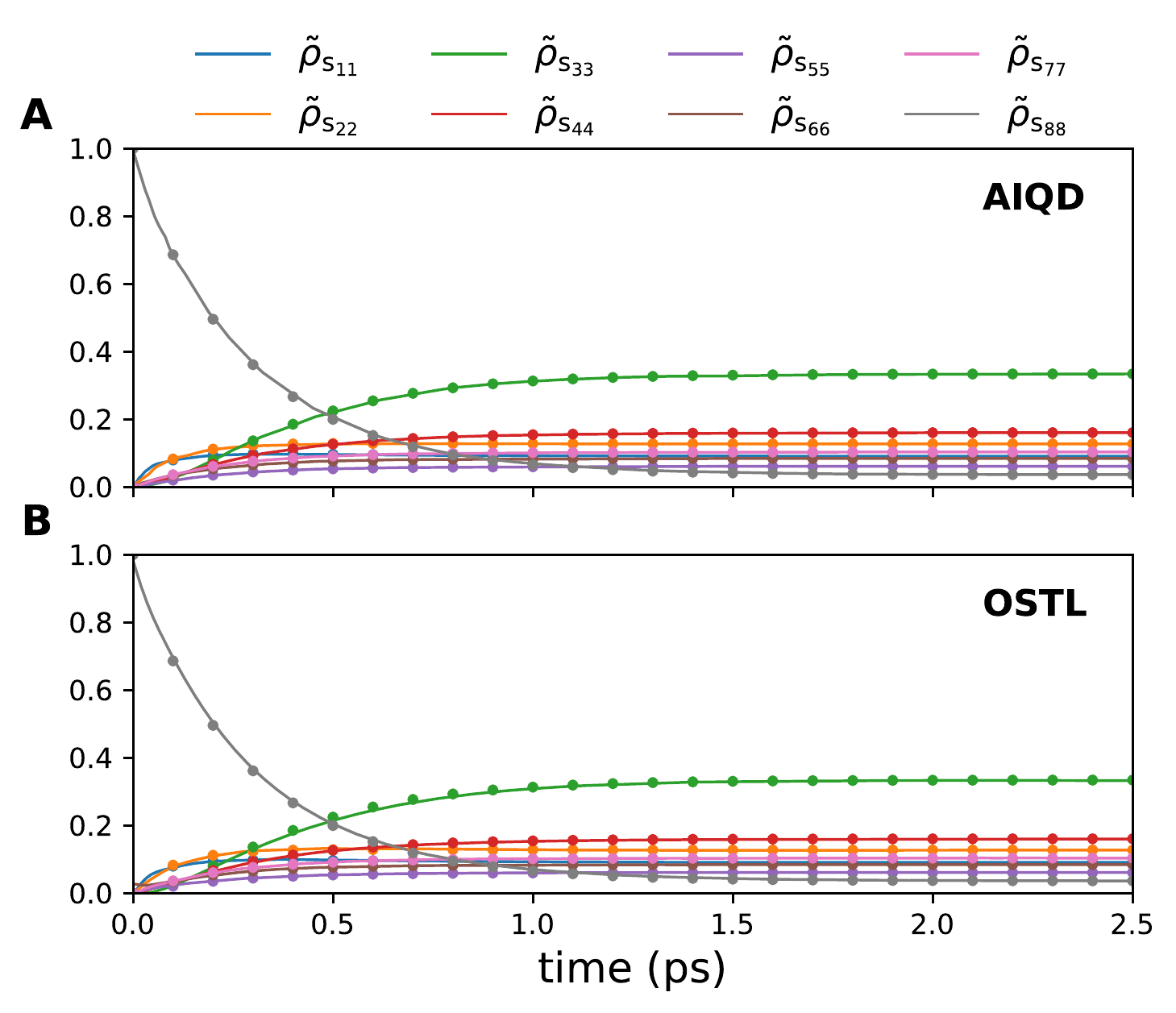}
    \caption{Time evolution of the diagonal elements of RDM $\tilde{\rho_{\rm s}}$ using (A) AIQD and (B) OSTL approaches. The initial excitation is on site 8 and other parameters are $\gamma = 200$~cm$^{-1}$, $\lambda =130$~cm$^{-1}$, $T=330$~K. The results are compared to the reference LTLME method (dots). For the time evolution of the prominent off-diagonal elements, the readers are referred to Fig.~\ref{fig:site8_coherence}.}
    \label{fig:site8_pop}
\end{suppfigure}
\begin{suppfigure}
    \centering
    \includegraphics[width=\textwidth]{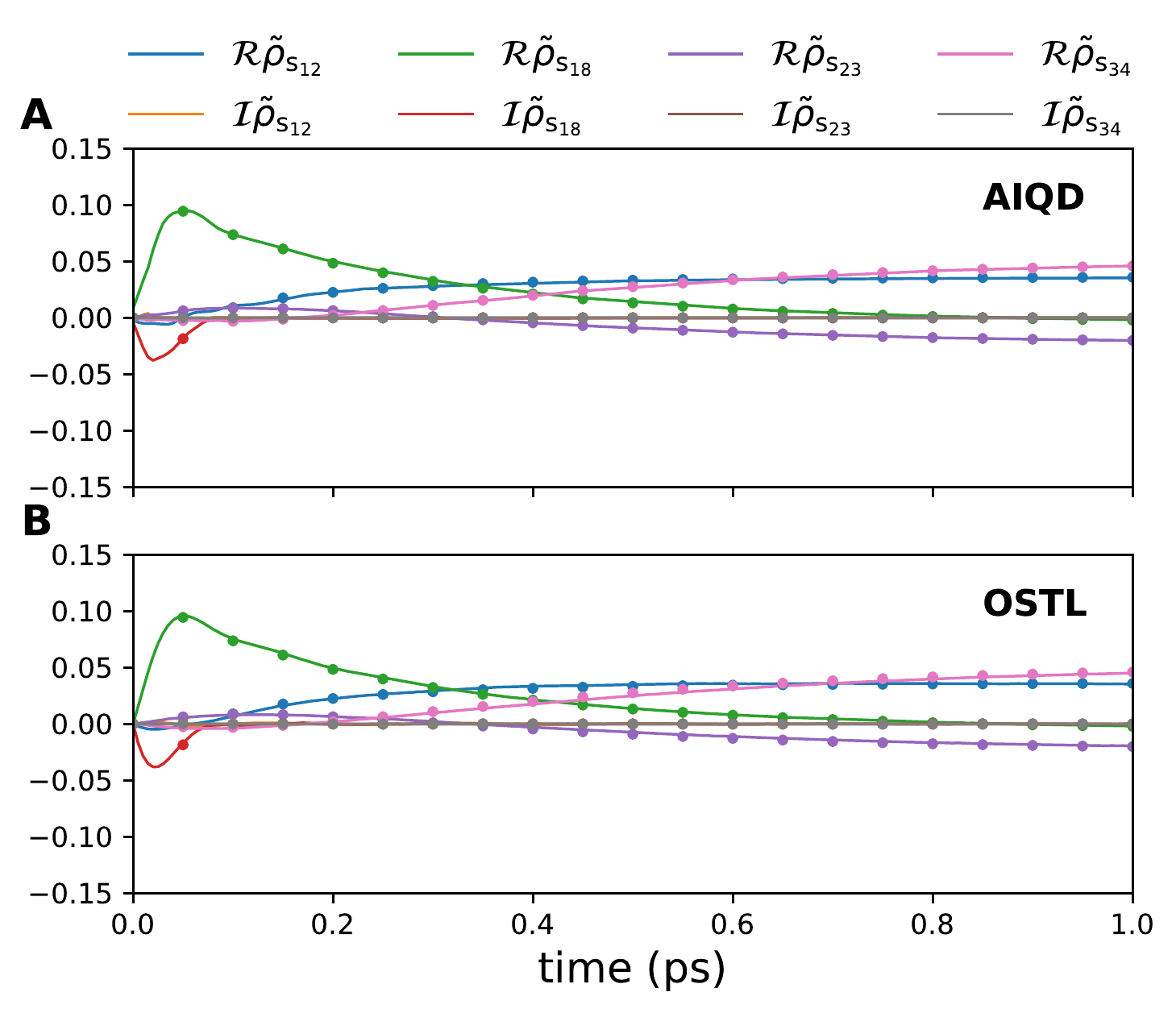}
    \caption{The time evolution of the dominant off-diagonal terms (i.e., $\tilde{\rho}_{\rm s_{mn}}, m\neq n$) using (A) AIQD and (B) OSTL approaches. The initial excitation is considered on site 8. The calligraphic $R$ and $I$ represent the real and imaginary parts, respectively. The results are compared to the reference LTLME method (dots). The time evolution of diagonal terms and the corresponding simulation parameters are given in Fig.~\ref{fig:site8_pop}.}
    \label{fig:site8_coherence}
\end{suppfigure}